# One-Dimensional Nature of Pairing and Superconductivity at the LaAlO$_3$/SrTiO$_3$ Interface

Yun-Yi Pai,*,† Hyungwoo Lee,‡ Jung-Woo Lee,‡ Anil Annadi, Guanglei Cheng,*,† Shicheng Lu,*,† Michelle Tomczyk,*,† Mengchen Huang,*,† Chang-Beom Eom,‡ Patrick Irvin,*,† Jeremy Levy,*,†

*Department of Physics and Astronomy, University of Pittsburgh, Pittsburgh, PA 15260, USA

†Pittsburgh Quantum Institute, Pittsburgh, PA, 15260 USA

‡Department of Materials Science and Engineering, University of Wisconsin-Madison, Madison, WI 53706, USA

**We examine superconductivity in LaAlO$_3$/SrTiO$_3$ channels in which the channel width transitions from the 1D to 2D regime. The superconducting critical current is independent of the channel width and increases approximately linearly with the number of parallel channels. Signatures of electron pairing outside of the superconducting phase are also independent of channel width. Collectively, these results indicate that electron pairing and superconductivity exist at the boundary of these channels and are absent within the interior region of the channels. The intrinsic 1D nature of superconductivity at the LaAlO$_3$/SrTiO$_3$ interface imposes strong physical constraints on possible electron pairing mechanisms.**

Strontium titanate (SrTiO$_3$ or STO) is a superconducting semiconductor [1] whose pairing mechanism has remained unresolved for more than half of a century. Its behavior is similar to that of high-temperature superconductors in many superficial aspects: both exhibit a dome-shaped superconducting transition temperature versus doping concentration [2], a low-density pseudogap phase [3], a small Fermi energy compared to the Debye frequency [4], and proximity to additional phase transitions [5,6]. that A wide range of pairing mechanisms responsible for superconductivity have been considered, including longitudinal optical phonons [7-9], antiferrodistortive modes [10], ferroelectric modes [11], plasmons [12], plasmons with optical phonons [13], and Jahn-Teller bipolarons [14]. Recently, interest in the superconducting properties of STO was revived by the development of STO-based heterostructures, and in the LaAlO$_3$/SrTiO$_3$ (LAO/STO) system [15] in particular. The two-dimensional interface supports superconductivity [16], and it can be electrostatically gated to trace out a superconducting dome [17], similar to the dome originally obtained through chemical doping [2].

Further reduction in dimensionality has become possible through the use of conductive-atomic force microscope (c-AFM) lithography [18,19], which relies on AFM tip-controlled protonation/deprotonation [20,21] of the LAO surface. A variety of quasi-1D and confined ("quasi-0D") structures have been created, including superconducting nanowires [22], ballistic 1D channels [23], and single-electron transistors [24], that revealed the existence of electron pairing outside the superconducting state [25]. Despite all of the new information about the superconducting phase, the origin of the pairing "glue" remains a mystery.

Here, we systematically investigate low-temperature transport behavior in conducting channels, formed at the LAO/STO interface, with widths ranging between 10 nm and 1 μm. LAO/STO heterostructures are grown by pulsed laser deposition with growth parameters reported in Ref. [21]. The thickness of LAO is fixed to 3.4 u.c., close to the metal-insulator transition [26]. Electrical contact to the LAO/STO interface is made by Ar$^+$ etching (25 nm) followed by sputter depositing Ti/Au (4 nm/25 nm). Conductive nanostructures at the LAO/STO interface are subsequently created using c-AFM lithography [18].

The first family of devices considered here (FIG. 1(a)) consists of three sections in series with characteristic widths $w_1 \sim 10$ nm, $w_2 = 100$ nm, and $w_3 = 1$ μm. All three sections (which are subsequently referred to as $w_1$, $w_2$, and $w_3$ sections) have the same length $L= 3$ μm. The $w_1$ section is created by writing a single line, while sections $w_2$ and $w_3$ are created by raster-scanning a rectangular area along the two principal axes. Conductive rectangular shapes separate the individual wire segments, enabling each to be monitored simultaneously and independently.

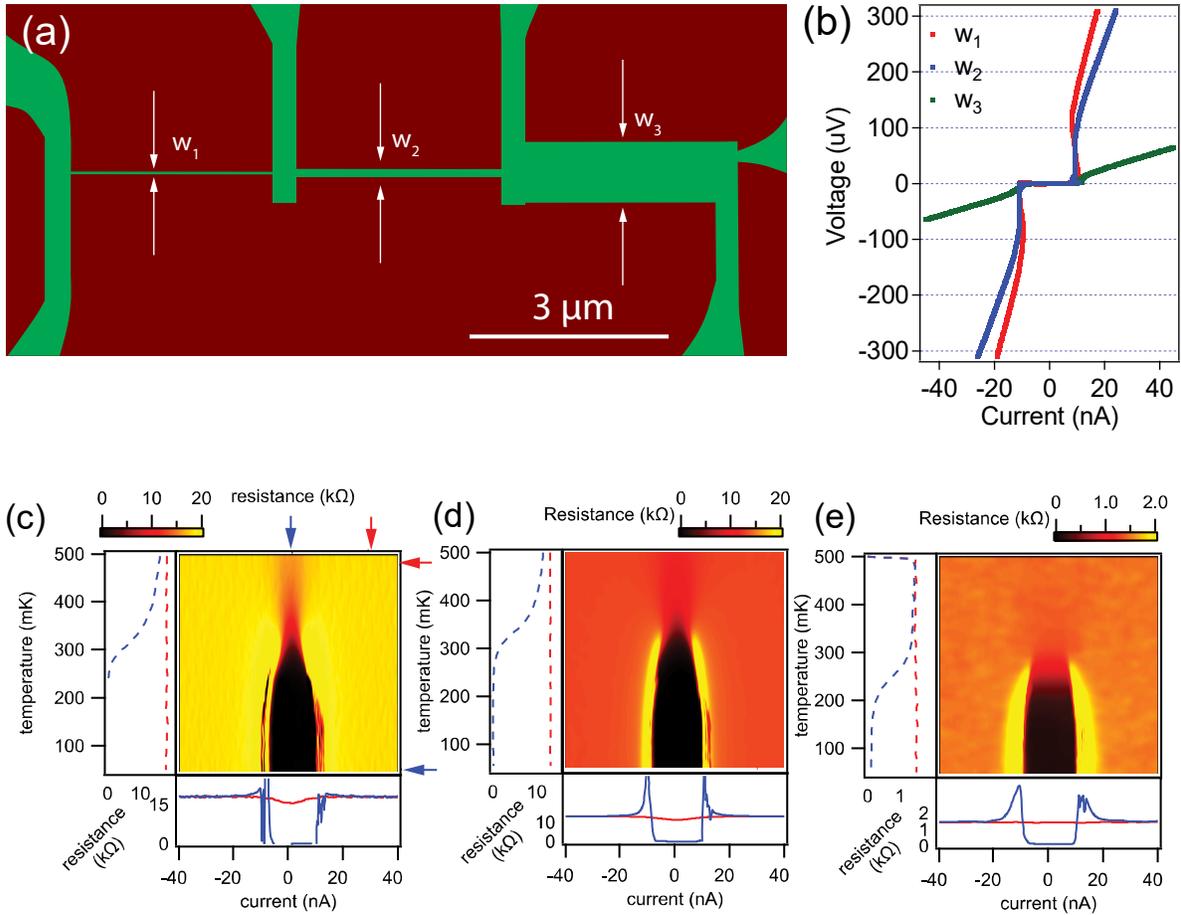

FIG. 1. (a) LAO/STO device (top view) consisting of three sections with widths $w_1 = 10$ nm, $w_2 = 100$ nm, and $w_3 = 1000$ nm. All three sections have the same length $L = 3$ μm. Green (red) areas depict conducting (insulating) regions (b) *I-V* curves for different channels measured at 50 mK and $V_{bg}$ = -6.5 V. (c-e) Differential resistance ($dV/dI$) as a function of current and temperature. (c) $w_1$ device, (d) $w_2$ device, and (e) $w_3$ device. Data taken at $V_{bg}$ = -6.5 V and $B$=0 T.

After c-AFM lithography, the devices are transferred into a dilution refrigerator and cooled to a base temperature $T$~50 mK. Four-terminal current-voltage (*I-V*) measurements for each of the three sections are recorded as a function of out-of-plane magnetic field (*B*), temperature, and back-gate voltage ($V_{bg}$). We identify the sharp increase in differential resistance above a critical value $I_c$ with the superconducting switching current, which provides a lower bound for the actual critical current due to various phase-slip mechanisms [27]. The results reported here are representative of three nominally identical sets of devices that show qualitatively similar behavior. FIG. 1(b) shows the 4-terminal *I-V* curves for the three sections at a back-gate voltage of $V_g$ = -6.5 V and $T$ = 50 mK. While all three sections are superconducting, the critical current $I_{c,i}$ within each section is remarkably similar (~10 nA), i.e., *independent of the channel width*. By contrast, the normal-state resistance (*i.e.*, resistance under dc bias that exceeds $I_{c,i}$) decreases monotonically with increasing width: $R_1$=17 kΩ for $w_1$, $R_2$=11.5 kΩ for $w_2$, and $R_3$=1.4 KΩ for $w_3$. In particular, the resistance drop between $w_2$ and $w_3$ is nearly equal to the ratio of the widths $w_2/w_3$, indicating that the 1D-2D crossover takes place near 100 nm.

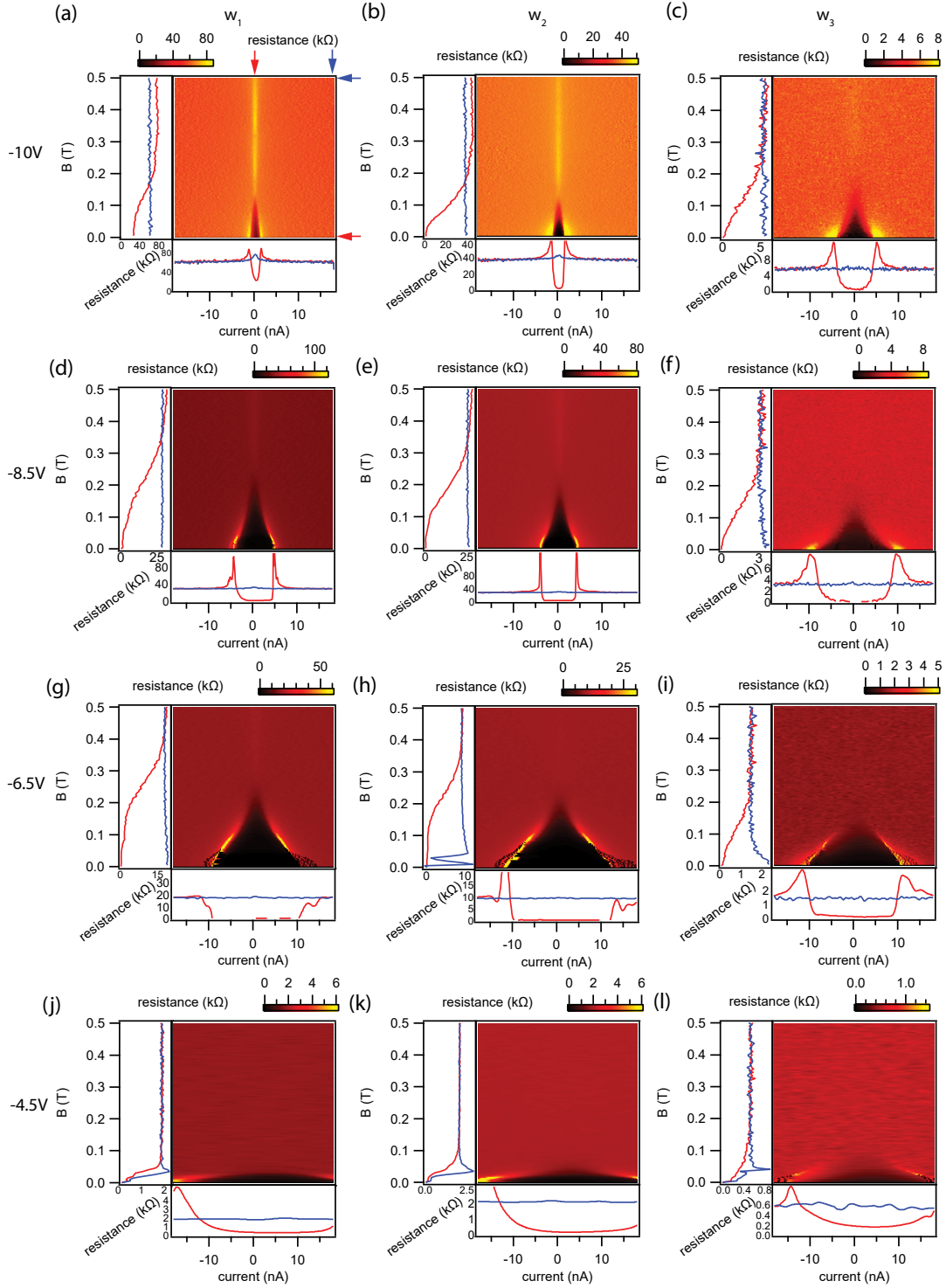

FIG. 2. Differential resistance ($dV/dI$), plotted as a function of current and magnetic field, for each of the three channels at different backgate voltages. The lower panel for each graph shows the horizontal linecuts at $B=0$ T (red) and $B=0.5$ T (blue). The left panel shows the vertical linecuts at bias current $I_1=0$ nA (red) and $I_2=14.5$ nA (blue).

Further insight into the superconducting nature of these channels comes from examining the differential resistance ($dV/dI$), obtained from numerical differentiation of the *I-V* curves. FIG. 1(c-e) shows the differential resistance of the three different sections as a function of current and temperature. Linecuts at fixed temperatures ($T_1 = 50$ mK, $T_2 = 475$ mK) and bias currents ($I_1 = 0$ nA, $I_2 = 300$ nA), indicated by arrows, are shown for each device. The superconducting transition temperature is about $T_c \sim 300$ mK for sections $w_1$ and $w_2$, and slightly lower ($T_c \sim 250$ mK) for section $w_3$. Notably, when $T > T_c$ a zero bias conductance peak is observed, for both $w_1$ and $w_2$ sections (FIG. 1(c,d)). This feature is much less pronounced for the widest section, $w_3$ (FIG. 1(e)).

FIG. 2 tracks the differential resistance of the three sections as a function of bias current, magnetic field and gate voltage. Intensity plots of $dV/dI(B, I)$ are shown for selected values of $V_{bg}$. A number of observations can be made:

(i) The superconducting upper critical field $\mu_0 H_{c2}$ initially increases with increasing backgate and then decreases. This non-monotononic dependence is reminiscent of the superconducting dome commonly observed for this interface.

(ii) The critical current increases monotonically when increasing the backgate voltages for all three sections. The critical currents for the three sections are strikingly similar to one another, except at the most negative backgate value.

(iii) A zero-bias conductance dip is observed above $H_{c2}$ (e.g., FIG. 2 (a)) and is most pronounced at the most negative backgate voltages. The conductance dip for the three channel widths (FIG. 3) is nearly the same for the $w_1$ and $w_2$ channels, and it is approximately twice as large for the $w_3$ channel.

(iv) The pronounced differences between the widest section, $w_3$, and the two narrower sections cannot be ascribed to the writing process, since section $w_2$ is created by raster-scanning and $w_1$ is created by moving the AFM tip along a single line. In other words, the fact that sections $w_1$ and $w_2$ behave similarly, and significantly different from section $w_3$, illustrates that the pairing is influenced by the physical geometry rather than the method in which the conducting regions are produced.

The results presented thus far are consistent with a scenario in which pairing and superconductivity exists within a quasi-1D ($w$~50-100 nm) portion of the channel, and in which pairing and superconductivity coexists with a parallel, non-superconducting (2D) bulk phase. The superconducting critical current density for section $w_3$ ($j_C$ ~ 10 nA/μm) is comparable to what has been reported for the bulk LAO/STO interface [16,28], while the critical current density of section $w_2$ is an order of magnitude higher .

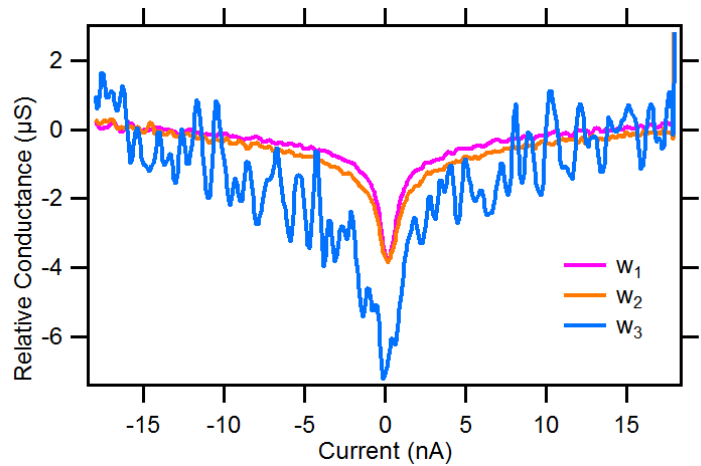

**FIG. 3.** The zero-bias conductance drop for the three sections. The curves are a result of averaging over the range of magnetic field values $B$=0.23 T—0.73 T Note that the size of the zero-bias conductance dip is similar for sections $w_1$ and $w_2$, and approximately twice as large for section $w_3$.

A possible explanation is that superconductivity exists only within a 1D region of the 2D channel, i.e., the outer edge(s). To test the hypothesis, we investigate a second type of device, as shown in Fig. 5(a). The device consists of three sections in series: from left, a single $w_1$ = 10 nm nanowire ("$w_1$"), a section of five parallel nanowires of width 10 nm ("$5w_1$") spaced 200 nm apart, and a $w_3$=1 μm section ("$w_3$"). The differential resistance $dV/dI$ ($I$, $V_{sg}$) is measured for each of the three sections (Figs. 5 (b-d)). The critical current for the $5w_1$ channel is 4-5 times larger than the other two sections, and it exhibits a different dependence on back-gate voltage. Meanwhile, channels $w_1$ and $w_3$ have similar superconducting critical currents; however, $w_3$ possesses a non-superconducting, parallel conductance that is an order-of-magnitude larger than channel $w_1$. This second class of experiments support the idea that superconductivity is associated with the channel boundaries, and that the interior bulk of the channels do not form a superconducting phase.

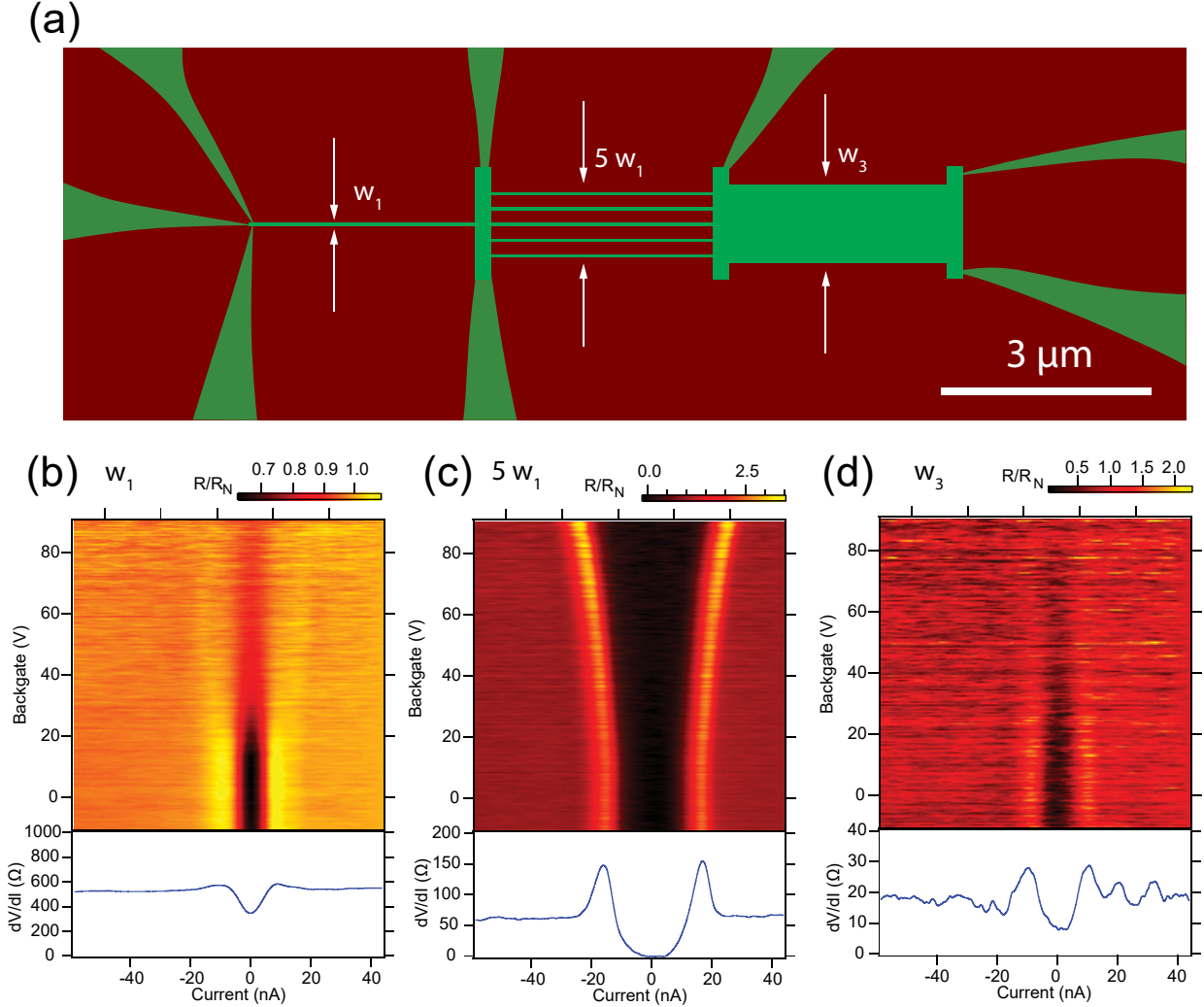

FIG. 4. (a) Multi-channel device consist of three sections. Left: single nanowire ($w_1$). Middle: five nanowires in parallel, 200 nm apart ($5w_1$). Right: 1 μm-wide channel ($w_3$). All three sections have the same length $L = 3$ μm. (b) The normalized differential resistance as a function of current and the backgate voltage, for the single nanowire section. The differential resistance in the color scale is normalized with respect to the normal state value. Lower panel: linecut of the raw differential resistance for the $w_1$ section, (c) $5w_1$ section, and (d) $w_3$ section, at backgate $V = 0$ V.

What might cause only the conducting boundaries of these channels to be superconducting? One possibility is that the center of the conductive channels is overdoped, i.e., on the high-density side of the superconducting dome, while the surrounding area is insulating, i.e., underdoped. In this scenario, a quasi-1D strip for which the doping is optimal should exist along each boundary (FIG. 5(a)). This simple picture satisfactorily predicts a width-independent critical current, and gives the correct scaling of parallel background conductance. Unanswered in this scenario is the question of why there should be a superconducting dome in the first place.

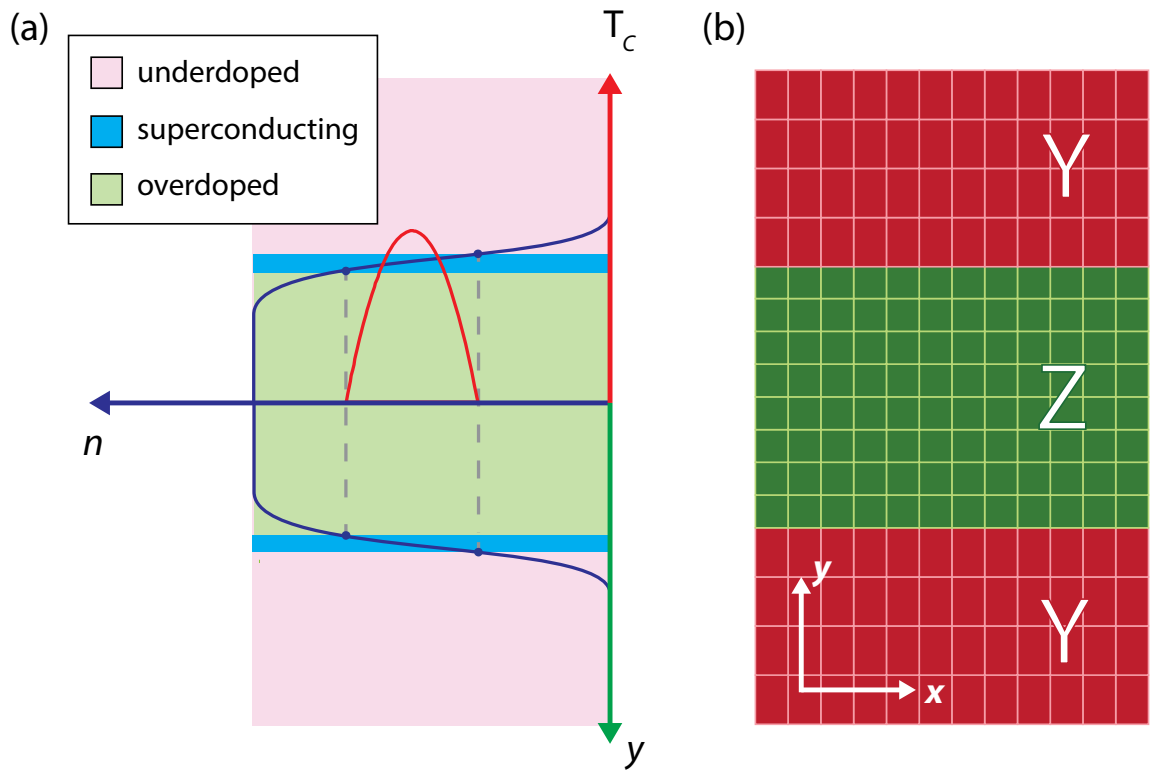

FIG. 5. (a) Schematic depicting quasi-1D region of optimal doping near the boundary of a 2D conducting channel. Dashed lines indicate lower and upper boundaries of the superconducting dome (red curve). Electron density profile (blue curve) traverses the entire superconducting dome, leaving a narrow region of critical doping. (b) Possible ferroelastic domain structure associated with conductive channel. Ferroelastic domain walls coincide with conductive boundaries.

STO undergoes a cubic-to-tetragonal antiferrodistortive transition at $T_{AFD}$ = 105 K. The transition combines antiphase rotations of $TiO_6$ cages with elongation of the unit call along the axis of the rotation. Below this transition, ferroelastic domains form with different orientations (*X, Y, Z*), separated by nanometer-scale domain walls. These domain walls can be driven by electrostatic gating [29], and are observed to be highly conductive [30]. Piezoelectric force microscopy imaging experiments on conductive LAO/STO nanostructures show that conductive regions formed by c-AFM lithography form *Z*-oriented tetragonal domains, even at room temperature [31]. This domain configuration is expected to persist to low temperatures, surrounded by regions that have strain-compensating X or Y domains (FIG. 5(b)). Ferroelastic domain boundaries thus naturally coincide with the edges that separate conducting and insulating regions.

If ferroelastic domain walls indeed bracket the edges of conducting nanostructures, one may naturally wonder whether they can mediate electron pairing. Ferroelastic domain walls possess structural, electronic, and point-defect properties that differ significantly from the uniform regions. The domain walls may fluctuate dynamically and couple to electronic states, yielding an attractive interaction. Alternatively, ferroelastic domain walls may trap high densities of oxygen vacancies [32] or other point defects that act as negative-U centers [33,34].

The presence and relevance of quasi-1D channels is not restricted to the artificially constructed channels created by c-AFM lithography. A variety of spatially-resolved imaging techniques have revealed strongly inhomogeneous electron transport at the 2D LAO/STO interface, and have demonstrated that current flows preferentially along ferroelastic domain boundaries, affecting properties in both the normal state [30,35,36] and superconducting [37] regime.

Regardless of the pairing mechanism, superconductivity in the strict 1D limit is of fundamental interest of its own right [27,38]. Low-dimensional superconductivity has been considered in several proposals to support Majorana fermions, which have not been observed in this system so far.

In conclusion, we have presented evidence that electron pairing and superconductivity at the LAO/STO interface naturally exists within quasi-1D channels at the edge of conducting 2D regions. The conclusion is supported by transport measurements for two families of devices in which the size and number of nanoscale channels is systematically varied. These experimental results provide stringent constraints on the microscopic mechanism of electron pairing and superconductivity in STO.

AUTHOR INFORMATION

Corresponding Author

*E-mail: jlevy@pitt.edu

NOTES

The authors declare no competing financial interest.

ACKNOWLEDGMENTS

We acknowledge helpful discussions with David Pekker and Anthony Tylan-Tyler. J.L. acknowledges support from the Vannevar Bush Faculty Fellowship program sponsored by the Basic Research Office of the Assistant Secretary of Defense for Research and Engineering and funded by the Office of Naval Research (N00014-15-1-2847).

REFERENCES

[1] J. F. Schooley, W. R. Hosler, and M. L. Cohen, Phys Rev Lett **12**, 474 (1964).
[2] C. S. Koonce, M. L. Cohen, J. F. Schooley, W. R. Hosler, and E. R. Pfeiffer, Phys Rev **163**, 380 (1967).
[3] C. Richter *et al.*, Nature **502**, 528 (2013).
[4] C. Lin and A. A. Demkov, Phys Rev Lett **111**, 217601 (2013).
[5] S. E. Rowley, L. J. Spalek, R. P. Smith, M. P. M. Dean, M. Itoh, J. F. Scott, G. G. Lonzarich, and S. S. Saxena, Nat Phys **10**, 367 (2014).
[6] C. W. Rischau *et al.*, Nat Phys **advance online publication** (2017).
[7] M. L. Cohen, Reviews of Modern Physics **36**, 240 (1964).
[8] A. Baratoff and G. Binnig, Physica B+C **108**, 1335 (1981).
[9] L. P. Gor'kov, Proceedings of the National Academy of Sciences **113**, 4646 (2016).
[10] J. Appel, Phys Rev **180**, 508 (1969).
[11] J. M. Edge, Y. Kedem, U. Aschauer, N. A. Spaldin, and A. V. Balatsky, Phys Rev Lett **115**, 247002 (2015).
[12] J. Ruhman and P. A. Lee, Phys Rev B **94**, 224515 (2016).

[13] Y. Takada, Journal of the Physical Society of Japan **49**, 1267 (1980).
[14] A. Stashans, H. Pinto, and P. Sanchez, J Low Temp Phys **130**, 415 (2003).
[15] A. Ohtomo and H. Y. Hwang, Nature **427**, 423 (2004).
[16] N. Reyren *et al.*, Science **317**, 1196 (2007).
[17] A. D. Caviglia *et al.*, Nature **456**, 624 (2008).
[18] C. Cen, S. Thiel, G. Hammerl, C. W. Schneider, K. E. Andersen, C. S. Hellberg, J. Mannhart, and J. Levy, Nature Materials **7**, 298, 10.1038/nmat2136 (2008).
[19] C. Cen, S. Thiel, J. Mannhart, and J. Levy, Science **323**, 1026 (2009).
[20] F. Bi, D. F. Bogorin, C. Cen, C. W. Bark, J. W. Park, C. B. Eom, and J. Levy, Applied Physics Letters **97**, 173110 (2010).
[21] K. A. Brown *et al.*, Nature Communications **7**, 10681 (2016).
[22] J. P. Veazey *et al.*, Nanotechnology **24**, 375201 (2013).
[23] M. Tomczyk *et al.*, Phys Rev Lett **117**, 096801 (2016).
[24] G. Cheng *et al.*, Nat Nanotechnol **6**, 343 (2011).
[25] G. Cheng *et al.*, Nature **521**, 196 (2015).
[26] S. Thiel, G. Hammerl, A. Schmehl, C. W. Schneider, and J. Mannhart, Science **313**, 1942 (2006).
[27] F. Altomare and A. M. Chang, *One-Dimensional Superconductivity in Nanowires* (Wiley, 2013).
[28] S. Hurand *et al.*, Scientific Reports **5**, 12751 (2015).
[29] M. Honig, J. A. Sulpizio, J. Drori, A. Joshua, E. Zeldov, and S. Ilani, Nature Materials **12**, 1112 (2013).
[30] B. Kalisky *et al.*, Nature Materials **12**, 1091 (2013).
[31] M. Huang, F. Bi, S. Ryu, C.-B. Eom, P. Irvin, and J. Levy, APL Materials **1**, 052110 (2013).
[32] L. Goncalves-Ferreira, S. A. T. Redfern, E. Artacho, E. Salje, and W. T. Lee, Phys Rev B **81**, 024109 (2010).
[33] P. W. Anderson, Phys Rev Lett **34**, 953 (1975).
[34] Y. Matsushita, H. Bluhm, T. H. Geballe, and I. R. Fisher, Phys Rev Lett **94**, 157002 (2005).
[35] Y. Frenkel, N. Haham, Y. Shperber, C. Bell, Y. Xie, Z. Chen, Y. Hikita, H. Y. Hwang, and B. Kalisky, ACS Applied Materials & Interfaces **8**, 12514 (2016).
[36] N. J. Goble, R. Akrobetu, H. Zaid, S. Sucharitakul, M.-H. Berger, A. Sehirlioglu, and X. P. A. Gao, Scientific Reports **7**, 44361 (2017).
[37] H. Noad *et al.*, Phys Rev B **94**, 174516 (2016).
[38] A. Bezryadin, *Superconductivity in Nanowires: Fabrication and Quantum Transport* (Wiley, 2013).